\documentclass[prl,twocolumn]{revtex4}%
\usepackage{amsfonts}
\usepackage{amsmath}
\usepackage{amssymb}
\usepackage{graphicx}

\begin{document}
\title{Symmetry Breaking in Induced-Charge Electrophoresis}
\author{Ehud Yariv}
\email{yarive@technion.ac.il}
 \affiliation{Faculty of Mechanical Engineering, Technion --- Israel Institute of Technology, Haifa 32000, Isreal}
\begin{abstract}
The electrophoretic motion of a conducting particle, driven by an induced-charge mechanism, is analyzed. The dependence of the motion upon particle shape
 is embodied in four tensorial coefficients  that relate the particle velocities to the externally-applied electric field. Several families of particle shapes, whose members are unaffected by the  field, are identified via use
of symmetry arguments. Other particles translate and/or rotate in response to the imposed field, even if their net electric charge vanishes. The coefficients are represented as surface integrals of the electric potential over the particle boundary, thereby  eliminating the need to solve the flow field.  
\end{abstract}
\maketitle
Traditionally, the term electrophoresis applies to the motion of colloidal
particles 
through an electrolyte
solution. This motion results from the interaction between an applied
electric field, say $\mathbf{E}_{\infty }$, and the net ionic charge
accommodated in the Debye cloud surrounding the particle boundary (this
charge being equal in magnitude and opposite in sign to that adsorbed at the
particle surface). In many situations, the Debye-layer thickness is small
compared with particle size. The electrokinetic flow is then
described by an approximated equation set, obtained via use of singular
perturbations \cite{OBrien83}. In that asymptotic description, the Debye screening length
assumes zero size. Thus, the electric potential $\varphi $ and velocity
field $\mathbf{v}$ in the electrically-neutral fluid domain are respectively
governed by Laplace's and Stokes' equations, whereas the presence of of the
layer is reflected by the boundary conditions that apply at the particle
boundary $\mathcal{S}$. These consist of  the
no-flux condition, $\mathbf{\hat{n}}\left( \mathbf{x}_{s}\right) \cdot 
\mathbf{E}\left( \mathbf{x}_{s}\right) =0$, governing the normal component
of the electric field, and the slip condition, $\mathbf{v}_{\mathrm{S}%
}\left( \mathbf{x}_{s}\right) =-\left( \varepsilon /\mu \right) \zeta \left( 
\mathbf{x}_{s}\right) \mathbf{E}\left( \mathbf{x}_{s}\right)$, governing
the relative fluid--particle motion. These conditions
respectively reflect the particle impermeability  to ionic current and
the net action of the electric field on the charged Debye layer. Here, $\mathbf{\hat{n}}$ is an outward-pointing unit vector normal to $%
\mathcal{S}$, $\mathbf{x}_{s}$ is a position vector located on $\mathcal{S}$, $%
\varepsilon $ and $\mu $ respectively denote the permittivity and viscosity
of the solution, the ``zeta potential" $\zeta $ constitutes the
local value of the potential jump across the Debye layer, and $\mathbf{E}%
=-\nabla \varphi $ is the electric field.

The preceding description is adequate for dielectric media,  capable of
carrying bound electric charge. Electrokinetic flow can also be generated about a conducting body; such flow, however, is driven by a polarization mechanism, made possible by the  charge mobility. As in the dielectric case, the electric charge is 
also situated on the surface; this, however, is not the result of any
physicochemical bonding, 
but rather a simple consequence of Gauss law \cite{footnote}. 
Moreover, the surface charge in a
conductor distributes over its surface so as to guarantee zero electric
field in its interior. The resulting dependence of the zeta potential
upon $\mathbf{E}_{\infty }$ gives rise to a nonlinear slip mechanism, which
is responsible for a rich variety of physical phenomena. 
 The most notable feature in this induced
electrokinetic effect is the generation of electrokinetic flow about an
initially-uncharged particle, this flow having no counterpart in
conventional linear electrophoresis.

Flows driven by this induced mechanism, addressing both DC and AC external fields, have been discussed mainly in the
Ukrainian literature (see~\cite{Murtsovkin1996}). 
A detailed description of induced-charge electrokinetics
appeared in a recent paper by
Squires and Bazant \cite{SquiresBazant2004}. The authors, discussing a broader
context of flows about polarizable media, coined the terms
``induced-charge electro-osmosis'' (ICEO) and ``induced-charge electrophoresis'' (ICEP) to respectively describe the associated fluid and particle motion.
They also suggested 
\cite{BazantSquires2004} that the combination of nonlinear slip with
symmetry breaking can lead to useful configurations which can perform a
variety of microfluidic operations (such as pumping and mixing). Their suggestions focus upon electro-osmotic flows relative to fixed conducting
elements (which may be held at a constant electric potential), rather then
 electrophoretic motion of freely-suspended particles, which is addressed
herein. Nevertheless, electrophoretic analyses tend to provides insight to
the inverse problem of electro-osmosis.

The simplest ICEO situation is that of an initially
uncharged conducting particle which is positioned in a uniform and constant applied
field $\mathbf{E}_{\infty }$. During a short time interval, a 
Faradaic current charges the Debye layer in proximity to the particle surface,
with an equal and opposite charge
being set up on the  surface itself. The induced potential difference
across the layer, namely the zeta potential, is proportional to $\mathbf{E}%
_{\infty }$. Accordingly, the electrokinetic slip that occurs at $\mathcal{S}
$\ is quadratic in $\mathbf{E}_{\infty }$. Given the linearity of the flow
problem, so are all the variables describing the resulting fluid (and
perhaps particle) motion. 

The prototypical ICEO problem, which was investigated in \cite%
{Levich1962}\ as a model for the electrokinetic flow about mercury drops,
 entails a spherical particle. Owing to the fore-aft symmetry of this
 flow \cite{Murtsovkin1996}, no
hydrodynamic force (or torque) is exerted on the particle.
This dipolar symmetry is obviously absent in flows about
 more general particle shapes. In principle,
 any deviation from a spherical  shape may lead to
hydrodynamic force and torque, with a consequent 
particle motion. Indeed, such motion has been experimentally observed
\cite{GamayunovMurtsovkin1987}. 

The purpose of this paper is to elucidate the interesting
properties of this nonlinear phenomenon. 
I consider here a relatively simple case,
addressing the electrophoretic motion of a freely-suspended conducting
particle in a uniform and constant externally-applied field. (The present results are easily generalized to the case of AC fields.) 
The analysis  begins with the case of an initially-uncharged particle
and is followed by discussion of non-zero net charge  effects.

Consider an uncharged conducting particle of an arbitrary shape and of
linear dimension $a$, which is positioned in an electrolyte solution in the
presence of an externally-applied constant electric field, $\mathbf{E}%
_{\infty }=E_{\infty }\mathbf{\hat{E}}$ ($\mathbf{\hat{E}}$\ being a unit
vector in the direction of $\mathbf{E}_{\infty }$). When describing the
governing field equations, the position vector is conveniently measured from
an (arbitrary) point $O$ attached to the particle. It is also convenient to
describe the particle motion by the instantaneous velocity of point $O$, $%
\mathbf{U}_{O}$, together with its angular velocity $\boldsymbol{\Omega }$.

For conducting particles which are impermeable to the current carrying ions
(as is usually the case) the no-flux condition, $\mathbf{\hat{n}}\cdot \boldsymbol{\nabla }\varphi =0$, still holds. Thus, the electrostatic problem governing $\varphi $ consists of this condition, together with Laplace's equation in the
fluid domain, $\nabla ^{2}\varphi =0$, and the far field condition, $\boldsymbol{\nabla }\varphi \rightarrow -%
\mathbf{E}_{\infty }.$ This problem (which also describes the field about
 a comparable dielectric particle) is linear and homogeneous in $\varphi $
and $\mathbf{E}_{\infty }$, from which it follows that $\varphi $ must be a
linear function of $\mathbf{E}_{\infty }$.
The attendant flow field is governed by the Stokes equations, the slip condition at $\mathcal{S}$, $%
\mathbf{v}=\mathbf{U}_{O}+\boldsymbol{\Omega }\times \mathbf{x}+\left(
\varepsilon \zeta /\mu \right) \boldsymbol{\nabla }\varphi $, and the
attenuation condition at large distances from $\mathcal{S}$, $\mathbf{v}%
\rightarrow \mathbf{0}$. For a freely
suspended particle, the requirement of vanishing hydrodynamic force and
torque completely determines the flow field, as well as the values of $%
\mathbf{U}_{O}$\ and $\boldsymbol{\Omega }$.

As in common in the literature (see~\cite{SquiresBazant2004}), it is assumed that the zeta potential is
sufficiently low to be proportional to the local surface charge density $q$, 
\begin{equation}
\label{linear capacitor}
\zeta =\frac{\lambda}{\varepsilon} q.
\end{equation}
Here, $\lambda $ is the Debye-layer thickness of the
Debye layer.
With no loss
of generality, the uniform value of the electric potential in the
particle interior is chosen as zero. Accordingly, the zeta potential is related to the potential value just outside the  layer, $\zeta =-\varphi|_\mathcal{S} $. 
It is the nonlinear slip term $\varphi\boldsymbol{\nabla }\varphi $ which is
responsible for the peculiar characteristics of ICEP.
Since the slip is the driver of the flow, this term  results in a
quadratic dependence of $\mathbf{v}$\ upon $\mathbf{E}_{\infty }$. Owing to
the linearity of the flow problem, the particle
translational and rotational velocities must possess the following
structure: 
\begin{subequations}
\label{def C D}
\begin{eqnarray}
\mathbf{U}_{O} &=&\frac{\varepsilon a}{\mu }\mathsf{C}_{O}:\mathbf{E}%
_{\infty }\mathbf{E}_{\infty },  \label{C def} \\
\mathbf{\Omega } &=&\frac{\varepsilon }{\mu }\mathsf{D}:\mathbf{E}_{\infty }%
\mathbf{E}_{\infty }.
\end{eqnarray}%
\end{subequations}
Here, $\mathsf{C}_{O}$\ is a dimensionless third-order tensor (which depends
upon the position of $O$) and $\mathsf{D}$\ is a dimensionless third-order
pseudo-tensor. These objects are intrinsic geometric properties of the
particle, independent of its size and  orientation relative to the imposed
field. In a sense, they constitute a lumped description of the electrophoretic
motion.

It is actually possible to demonstrate a variety of symmetry properties,
expressed through the coefficients $\mathsf{C}_{O}$\ and $\mathsf{D}$, for
various families of particle shapes. These properties may be
obtained without the need to explicitly solve the governing field equations.
The simplest example involves isotropic particles (such as spheres). For
that family, no vector (or tensor) can specify the particle geometry. Since
no isotropic third-order tensor exists \cite{Aris:62}, it is impossible to
construct a candidate for $\mathsf{C}_{O}$. Moreover, the only isotropic
third-order pseudo-tensor is $\boldsymbol{\epsilon }$, the alternating
triadic, which yields zero when contracted with the symmetric dyadic $%
\mathbf{E}_{\infty }\mathbf{E}_{\infty }$. Accordingly, isotropic particles
neither translate nor rotate.

\begin{figure}
\begin{center}
\label{Fig:geometry}
\includegraphics[scale=0.2]{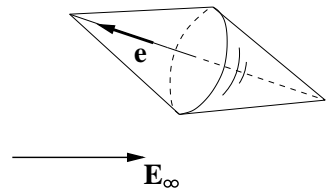}
\end{center}
\caption{An axisymmetric particle possessing fore-aft symmetry: in the absence of net charge, the particle remains stationary; in the presence of an initial charge, the particle translates but does not rotate.}
\end{figure}
A larger family is that of all axisymmetric particles possessing fore-aft
symmetries, such as spheroids and finite cylinders of circular cross section (see~Fig. 1). 
These particles are characterized by a single vector $\mathbf{e}$
attached to their axis of symmetry. The only third-order tensors which may
be constructed using $\mathbf{e}$\ are $\mathbf{eee}$ and various
permutations of $\mathsf{I}\mathbf{e}$ ($\mathsf{I}$ being the idem-factor).
These tensors, however, are odd functions of $\mathbf{e}$, and do not
satisfy the invariance to reversal of $\mathbf{e}$\ required from a body
possessing a fore-aft symmetry. Moreover, it is impossible to construct a
third-order pseudo-tensor using the single vector $\mathbf{e}$. Accordingly,
such particles behave like spheres: they 
neither translate nor rotate.

It appears that the generation of particle motion requires a stronger
symmetry breaking. Consider the case of axisymmetric particles lacking
fore-aft symmetry, such as cones and hemispheres. Again, these
particles are characterized by a single vector $\mathbf{e}$ attached to
their axis of symmetry, whence the only candidate for $\mathsf{D}$ is
again $\boldsymbol{\epsilon }$, which means that such particles do not
rotate. Both $\mathbf{U}_{O}$ and $\mathsf{C}_{O}$ are therefore independent
of $O$, and may be respectively denoted by $\mathbf{U}$\ and $\mathsf{C}$.
As with the previous family, the only candidates for $\mathsf{C}$ are $%
\mathbf{eee}$ and various permutations of $\mathsf{I}\mathbf{e}$. For
 bodies lacking fore-aft symmetry, $\mathsf{C}$ is not necessarily
invariant to reflection. Thus, such particles may translate (without
rotation) under the action of an external field.

In principle, both $\mathsf{C}_{O}$\ and $\mathsf{D}$\ are obtainable from the
detailed knowledge of the velocity field. In what follows, 
 integral expressions for these tensors are derived in terms of the electric potential
distribution. These expressions are calculated using the Lorentz reciprocal
theorem \cite{HappelBrenner:65}, thus avoiding the formidable  calculation
 of the
flow problem. For simplicity of presentation, this
procedure is demonstrated for non-rotating (e.g., axisymmetric) particles, 
for which the angular momentum balance is
automatically satisfied (the generalization to arbitrary particles is
trivial). In what follows, it is convenient to employ a dimensionless
notation, wherein length variables are normalized with $a$, velocities with $%
\varepsilon aE_{\infty }^{2}/\mu $, forces with $\varepsilon a^{2}E_{\infty
}^{2}$, stresses with $\varepsilon E_{\infty }^{2}$, and the electric potential
with $aE_{\infty }$.

Using this notation, the slip condition on $\mathcal{S}$ appears as $\mathbf{%
\widetilde{v}}=\mathbf{\widetilde{U}}-\widetilde{\varphi }\,\boldsymbol{%
\widetilde{\nabla }}\widetilde{\varphi }$ ($\widetilde{\cdots }$ denotes
here the dimensionless counterpart of the relevant physical quantity).
The linear flow problem is  decomposed
into two separate parts. The first describes pure translation of the
particle with a velocity $\mathbf{\widetilde{U}}$ (in the absence of any
slip); in the second  the particle is held fixed, and the flow
is generated by the imposed slip on $\mathcal{S}$, $\mathbf{\widetilde{v}}=$ $-\widetilde{%
\varphi }\,\boldsymbol{\widetilde{\nabla }}\widetilde{\varphi }$. In both
problems the velocity field satisfies the Stokes equations and attenuates at
large distances from the particle.

The force resulting from the first flow field is given by a 
 Stokes drag expression, $\mathbf{F}_{\mathrm{I}}=-\mathsf{K}\cdot \mathbf{%
\widetilde{U}}$, where the positive-definite dimensionless translation
tensor $\mathsf{K}$\ is an intrinsic property of the particle geometry \cite%
{HappelBrenner:65}. The force resulting from the second flow field is
provided by the following quadrature: 
\begin{equation}
\mathbf{F}_{\mathrm{II}}=-\oint_{\mathcal{S}}d\widetilde{A}\,\mathbf{\hat{n}}\cdot 
\boldsymbol{\Sigma }^{\dag }\cdot \,\boldsymbol{\widetilde{\nabla }}%
\widetilde{\varphi }
\end{equation}%
which follows from a variant of the reciprocal theorem \cite{Brenner:64}.
Here, $d\widetilde{A}$ is a dimensionless area element (normalized with $%
a^{2}$) and the (dimensionless) ``translational'' stress triadic field $\boldsymbol{\Sigma }$
(defined in \cite{HappelBrenner:65}) depends only upon the particle geometry
and the position vector of the fluid point on $\mathcal{S}$: explicitly, $%
\Sigma _{ijk}$ denotes the $ij$ component of the stress tensor that would
result from a pure translation of the particle with a unit velocity in the $k
$ direction. (The superscript $\dag $ signifies right transposition.)

The requirement of a force-free particle, $\mathbf{F}_{\mathrm{I}}+\mathbf{F}_{\mathrm{II}}=%
\mathbf{0}$, yields the following result: 
\begin{equation}
\mathbf{\widetilde{U}}=-\mathsf{K}^{-1}\cdot \oint_{\mathcal{S}}d\widetilde{A%
}\,\mathbf{\hat{n}}\cdot \boldsymbol{\Sigma }^{\dag }\cdot \widetilde{%
\varphi }\,\boldsymbol{\widetilde{\nabla }}\widetilde{\varphi }.
\label{intermediate rep}
\end{equation}%
Owing to the linearity of the electrostatic problem, it is clear that $%
\widetilde{\varphi }=\mathbf{\Phi }\left( \mathbf{\widetilde{x}}\right)
\cdot \mathbf{\hat{E}}$, where $\mathbf{\Phi }$, a dimensionless vector
function of position, can only depends upon  particle shape. Use of the
definition (\ref{C def}) furnishes the requisite representation,
\begin{equation}
\mathsf{C}=-\mathsf{K}^{-1}\cdot \oint_{\mathcal{S}}d\widetilde{A}\,\mathbf{%
\hat{n}}\cdot \boldsymbol{\Sigma }^{\dag }\cdot \boldsymbol{\widetilde{%
\nabla }}\mathbf{\Phi }\,\mathbf{\Phi }.  \label{c rep}
\end{equation}
It is  emphasize that this result holds only for a particle which does not possess
an initial charge. 

Consider now  a particle possessing an initial
charge $Q$ (or, equivalently, a net charge $Q$). In this situation,
 the zeta potential is not
quadratic in $\mathbf{E}_{\infty }$\ and the relations (\ref{def C D}) are
invalidated. Define an ``effective"
potential, $\psi\stackrel{\mathrm{def}}{=}Q\lambda /A\varepsilon $, where
 $A=\oint_{\mathcal{S}}dA$ is the particle
surface area. While it is customary to identify $\psi $ with the  zeta
potential [cf.~(\ref{linear capacitor})], I here avoid any such \textit{a priori} interpretation.

The flow about an initially-charged particle under the action of an applied field has been discussed in \cite%
{SquiresBazant2004}. Given the linear-capacitor approximation (\ref{linear capacitor}), additivity in charge implies additivity
in the zeta potential. Accordingly, the initially-charged  situation
was represented as a combination of two different flows: the first is
associated with a quadratic slip condition, and the second is that involved
in a linear
electrophoretic motion corresponding to a uniform zeta potential $\zeta=\psi $.
It is important to note, however, that the situation discussed in \cite%
{SquiresBazant2004} --- with a view towards microfluidic devices --- is
focused upon the specific case of a circular cylinder (a similar
representation is also valid for the case of spheres). Is that
interpretation valid for more general shapes?

The  additional scale $\psi $
modifies the translational and rotational relations (\ref{def C D}), 
which, following dimensional arguments, are replaced by:
\begin{subequations}
\begin{eqnarray}
\mathbf{U}_{O} &=&\frac{\varepsilon a}{\mu }\mathsf{A}_{O}:\mathbf{E}%
_{\infty }\mathbf{E}_{\infty }+\frac{\varepsilon }{\mu }\psi \mathsf{M}%
_{O}\cdot \mathbf{E}_{\infty }+\frac{\varepsilon }{\mu a}\psi ^{2}\mathbf{P}%
_{O},   \label{U with charge}  \\
\mathbf{\Omega } &=&\frac{\varepsilon }{\mu }\mathsf{B}:\mathbf{E}_{\infty }%
\mathbf{E}_{\infty }+\frac{\varepsilon }{\mu a}\psi \mathsf{N}\cdot \mathbf{E%
}_{\infty }+\frac{\varepsilon }{\mu a^{2}}\psi ^{2}\mathbf{R}. \label{Omega with charge}
\end{eqnarray}
\end{subequations}
The dimensionless objects $\mathsf{A}_{O}$, $\mathsf{M}_{O}$, and $\mathbf{P}%
_{O}$\ respectively denote a triadic, dyadic, and a vector, whereas the
dimensionless objects $\mathsf{B}$, $\mathsf{N}$, and $\mathbf{R}$
respectively denote a pseudo-triadic, pseudo-dyadic, and a pseudo-vector.
All six quantities are intrinsic geometric properties of the particle,
independent of the applied field and the value of $\psi $.

In the case $\mathbf{E}_{\infty }=\mathbf{0}$ the translational and
rotational velocities are respectively proportional to $\mathbf{P}_{O}$ and $%
\mathbf{R}$. Since the applied field is the driver of the motion, $%
\mathbf{P}_{O}$ and $\mathbf{R}$ must vanish in that case. These
quantities, however, are independent of $\mathbf{E}_{\infty }$, and must therefore vanish identically. Similar logic can be applied via the case $\psi =0$%
, which yields relations similar to (\ref{def C D}). Since $\mathsf{A}%
_{O}$ and $\mathsf{B}$ are independent of $\psi $, they must be respectively
identical to the coefficients $\mathsf{C}_{O}$\ and $\mathsf{D}$, which
describe the motion of particles possessing zero net charge.

The role of $\mathsf{M}_{O}$ and $\mathsf{N}$ is elucidated via the
asymptotic limit $aE_{\infty }/\psi \ll 1$, which corresponds to the 
 quadratic effect (associated with the action of $\mathbf{E}%
_{\infty }$ on charge induced by $\mathbf{E}_{\infty }$) being negligible
relative to the linear one (associated with the action of $\mathbf{E}%
_{\infty }$ on the initial charge distribution). In this case (\ref{U with charge})-(\ref{Omega with charge})
degenerate to:
\begin{subequations}
\label{mobilities}
\begin{eqnarray}
\mathbf{U}_{O} &=&\frac{\varepsilon }{\mu }\psi \mathsf{M}_{O}\cdot \mathbf{E%
}_{\infty },  \label{mobility U} \\
\mathbf{\Omega } &=&\frac{\varepsilon }{\mu a}\psi \mathsf{N}\cdot \mathbf{E}%
_{\infty }.
\end{eqnarray}%
These  are identical to the mobility relations governing
conventional electrophoresis (with no induced charge effects), except for an
 important difference which will be identified shortly. Since $\mathsf{M}%
_{O}$ and $\mathsf{N}$ are independent of $\psi $ and $\mathbf{E}_{\infty }$
(and , specifically, of the above-mentioned limit), they may be obtained using a linear electrophoretic analysis. It is therefore tempting to say (see~\cite{SquiresBazant2004})
that the mobility relations (\ref{mobilities}) are the same as those for a
linear electrophoresis of particles having a zeta potential $\psi $, which would imply that $\mathsf{M}_{O}$ and $%
\mathsf{N}$\ are governed by the Smoluchowski relation $\mathbf{U}=\left(
\varepsilon \psi /\mu \right) \mathbf{E}_{\infty }$ and $\mathbf{\Omega }=%
\mathbf{0}$. That would yield $\mathsf{M}=\mathsf{I}$ and $\mathsf{N}%
=\mathsf{0}$.

It is important, however, to emphasize that the linear electrophoretic
motion in the limit $aE_{\infty }/\psi \ll 1$ corresponds to a
charge distribution in a \emph{conducting} particle in the absence of any
imposed field. Only for the
case of a sphere (or a circular cylinder in two-dimensional flows) this distribution is uniform,
corresponding to a uniform value (namely $\psi $) of the zeta potential.
Thus, $\mathsf{M}_{O}$ and $\mathsf{N}$ would not be identical in general
to those predicted by the Smoluchowski relations. (Indeed, linear electrophoresis of non-uniformly charged dielectric particles \cite{Anderson1985} is qualitatively different
than that observed in the  uniform case.)  
It appears that 
the 
traditional focus on spherical and cylindrical bodies leads to a loss of
 physical effects which characterize nonlinear slip. As such, conclusions drawn from the analyses
of such specific shapes are rather restrictive and non-representative.

As in the case of  particles possessing no net charge, for which the relations (\ref{def C D}) apply, it is possible to deduce symmetry properties for the
motion (namely the values of $\mathsf{M}_{O}$ and $\mathsf{N}$) of initially-charged particles. With axisymmetric particles, for example, 
  $\mathsf{M}_{O}$ must be a
linear combination of $\mathsf{I}$ and $\mathbf{ee}$, whereas $\mathsf{N}$ must be
proportional to $\boldsymbol{\epsilon }\cdot \mathbf{e}$. In the low-field
limit, $aE_{\infty }/\psi \ll 1$, such particles do
not rotate, but in general may translate.

It is a simple matter to represent both $\mathsf{M}_{O}$ and $\mathsf{N}$\
as quadratures of the electric field, without the need to solve the flow
problem. As with the preceding analysis, which has led to the representation (%
\ref{c rep}), the present derivation focuses upon non-rotating (e.g., axisymmetric)
particles, for which $\mathsf{N}=\mathsf{0}$ and $\mathsf{M}_{O}$ is
independent of $O$ (again, the generalization to the general case is
trivial). In the limit $aE_{\infty }/\psi \ll 1$ the slip velocity
is given by $\left( \varepsilon \zeta /\mu \right) \boldsymbol{\nabla }%
\varphi $, with $\zeta $ being the zeta potential associated with the
initial charge distribution (which is a function of $\mathbf{x}_{s}$). 
It is convenient to use the previously employed
dimensionless notation, with which the present slip condition 
appears as $\mathbf{\widetilde{v}}=\mathbf{\widetilde{U}}+\left( \psi
/aE_{\infty }\right) \widetilde{\zeta }\,\boldsymbol{\widetilde{\nabla }}%
\widetilde{\varphi }$ (with $\widetilde{\zeta }=\zeta /\psi $). The
concomitant flow field is decomposed into two parts, the first describing
pure translation with a velocity $\mathbf{\widetilde{U}}$, 
and the second describing a stationary particle on which the velocity
slips with the prescribed value $\left( \psi /aE_{\infty }\right) \widetilde{%
\zeta }\,\boldsymbol{\widetilde{\nabla }}\widetilde{\varphi }$. A similar
procedure to that leading to (\ref{intermediate rep}) then  yields
\end{subequations}
\begin{equation}
\mathbf{\widetilde{U}}=\frac{\psi }{aE_{\infty }}\mathsf{K}^{-1}\cdot \oint_{%
\mathcal{S}}d\widetilde{A}\,\widetilde{\zeta }\,\mathbf{\hat{n}}\cdot 
\boldsymbol{\Sigma }^{\dag }\cdot \boldsymbol{\widetilde{\nabla }}\widetilde{%
\varphi }.
\end{equation}%
Use of the linear representation, $\widetilde{\varphi }=\mathbf{\Phi }\cdot 
\mathbf{\hat{E}}$, together with the definition (\ref{mobility U}) furnishes
the desired expression,%
\begin{equation}
\mathsf{M}=\mathsf{K}^{-1}\cdot \oint_{\mathcal{S}}d\widetilde{A}\,%
\widetilde{\zeta }\,\mathbf{\hat{n}}\cdot \boldsymbol{\Sigma }^{\dag }\cdot 
\boldsymbol{\widetilde{\nabla }}\mathbf{\Phi }.  \label{rep M}
\end{equation}%
It is re-emphasized that the zeta-potential distribution appearing in
(\ref{rep M}) corresponds to a surface charge arrangement on a \emph{%
conducting} particle in a vacuum (in the absence of any applied field),
whereas the electric field is identical to that about a \emph{dielectric} particle in a
vacuum.

For a sphere of radius $a$, the initial charge spreads uniformly, giving
rise to a charge density $Q/A$. Thus, $\psi $ is identical to the zeta
potential, and $\widetilde{\zeta }\equiv 1$. The potential distribution under the
action of an applied external field $\mathbf{E}_{\infty }$ is given by a
combination of a uniform field and a dipole, $\varphi =-\left(
1+a^{3}/2r^{3}\right) \mathbf{x}\cdot \mathbf{E}_{\infty }$, from which it
follows that $\boldsymbol{\widetilde{\nabla }}\mathbf{\Phi }|_{%
\mathcal{S}}=\left( 3/2\right) \left( \mathbf{\hat{n}\,\hat{n}}-\mathsf{I}%
\right) $. Also, from the solution of the Stokes equations for a
translating sphere \cite{HappelBrenner:65}, it is known that $\mathbf{\hat{n}%
}\cdot \boldsymbol{\Sigma }=-\left( 3/2\right) \mathsf{I}$ and $\mathsf{K}%
=6\pi \mathsf{I}$. Use of the integral relations $\oint_{\mathcal{S}}d%
\widetilde{A}=4\pi$ and $\oint_{\mathcal{S}}d\widetilde{A}\,%
\mathbf{\hat{n}\,\hat{n}}=\left( 4\pi /3\right) \mathsf{I}$\ readily yields
the Smoluchowski relation $\mathsf{M}=\mathsf{I}$. 
\begin{acknowledgments}
I thank  Martin Z. Bazant for turning my attention to ICEO, and for his valuable help in obtaining the Russian references. I thank Todd M. Squires for some useful discussions.
\end{acknowledgments}


\begin{thebibliography}{10}
\bibitem{footnote}
This would be true even in unsteady situations, the reason being the large
 ratio between typical conductivities of metallic 
conductors (say $6\times 10^{7}\mathrm{S/m%
}$ for copper) to those characterizing ionic solutions (where the conductivity, which depends
upon the ionic concentration, usually does not exceed $10^{-6}\mathrm{S/m}$).
\end{thebibliography}
\end{document}